%% file: main_aa.tex
\documentclass{aa}

\usepackage{graphicx}
\usepackage{graphics}
\usepackage{txfonts}
\usepackage{epsfig}
\usepackage{natbib}
\usepackage{hyperref}

\graphicspath{{./}{figures/}}
\newcommand{\ergs}{{\mbox{$\mathrm{erg\,s^{-1}}$}}}
\newcommand{\kms}{{\mbox{$\mathrm{km\,s^{-1}}$}}}

\begin{document}

\title{The resolved size and structure of hot dust\\ in the immediate vicinity of AGN}

\author{GRAVITY Collaboration:
  J.~Dexter\inst{1,23} 
\and J.~Shangguan\inst{1}
\and S.~H\"onig\inst{5}
\and M.~Kishimoto\inst{6} 
\and D.~Lutz\inst{1} 
\and H.~Netzer\inst{8}
\and R.~Davies\inst{1}
\and E.~Sturm\inst{1} 
\and O.~Pfuhl\inst{1,15}
\and A.~Amorim\inst{18,20}
\and M.~Baub\"{o}ck\inst{1}
\and W.~Brandner\inst{21}  
\and Y.~Cl\'enet\inst{2} 
\and P.~T.~de~Zeeuw\inst{1,16} 
\and A.~Eckart\inst{3,17} 
\and F.~Eisenhauer\inst{1} 
\and N.M.~F\"orster Schreiber\inst{1} 
\and F.~Gao\inst{1} 
\and P.~J.~V. ~Garcia\inst{14,19,20} 
\and R.~Genzel\inst{1,4} 
\and S.~Gillessen\inst{1} 
\and D.~Gratadour\inst{2}
\and A.~Jim\'{e}nez-Rosales\inst{1}
\and S.~Lacour\inst{2,1,15} 
\and F.~Millour\inst{7}  
\and T.~Ott\inst{1} 
\and T.~Paumard\inst{2} 
\and K.~Perraut\inst{12} 
\and G.~Perrin\inst{2} 
\and B.~M.~Peterson\inst{9,10,11} 
\and P.~O.~Petrucci\inst{12} 
\and M.~A.~Prieto\inst{22}
\and D.~Rouan\inst{2}
\and M.~Schartmann\inst{1}
\and T.~Shimizu\inst{1} 
\and A.~Sternberg\inst{8,13}
\and O.~Straub\inst{1}
\and C.~Straubmeier\inst{3} 
\and L.~J. Tacconi\inst{1} 
\and K.~Tristram\inst{14}  
\and P.~Vermot\inst{2} 
\and I.~Waisberg\inst{1,24}
\and F.~Widmann\inst{1} 
\and J.~Woillez\inst{15}}

\institute{
Max Planck Institute for Extraterrestrial Physics (MPE), Giessenbachstr.1, 85748 Garching, Germany
\and LESIA, Observatoire de Paris, Universit\'e PSL, CNRS, Sorbonne Universit\'e, Universit\'e de Paris, 5 place Jules Janssen, 92195 Meudon, France
\and I. Institute of Physics, University of Cologne, Z\"ulpicher Stra{\ss}e 77,50937 Cologne, Germany
\and Departments of Physics and Astronomy, Le Conte Hall, University of California, Berkeley, CA 94720, USA
\and Department of Physics and Astronomy, University of Southampton, Southampton, UK
\and Department of Astrophysics and Atmospheric Sciences, Kyoto Sangyo University, Japan
\and Universit\'e C\^ote d'Azur, Observatoire de la C\^ote d'Azur, CNRS, Laboratoire Lagrange, Nice, France
\and School of Physics and Astronomy, Tel Aviv University, Tel Aviv 69978, Israel
\and Department of Astronomy, The Ohio State University, Columbus, OH, USA
\and Center for Cosmology and AstroParticle Physics, The Ohio State University, Columbus, OH, USA
\and Space Telescope Science Institute, Baltimore, MD, USA
\and Univ. Grenoble Alpes, CNRS, IPAG, 38000 Grenoble, France
\and Center for Computational Astrophysics, Flatiron Institute, 162 5th Ave., New York, NY 10010, USA
\and European Southern Observatory, Casilla 19001, Santiago 19, Chile
\and European Southern Observatory, Karl-Schwarzschild-Str. 2, 85748 Garching, Germany
\and Sterrewacht Leiden, Leiden University, Postbus 9513, 2300 RA Leiden, The Netherlands
\and Max Planck Institute for Radio Astronomy, Bonn, Germany
\and Universidade de Lisboa - Faculdade de Ci\^{e}ncias, Campo Grande, 1749-016 Lisboa, Portugal
\and Faculdade de Engenharia, Universidade do Porto, rua Dr. Roberto Frias, 4200-465 Porto, Portugal
\and CENTRA - Centro de Astrof\'isica e Gravita\c{c}\~{a}o, IST, Universidade de Lisboa, 1049-001 Lisboa, Portugal
\and Max Planck Institute for Astronomy, K\"onigstuhl 17, 69117, Heidelberg, Germany
\and Instituto de Astrof\'isica de Canarias (IAC), E-38200 La Laguna,
Tenerife, Spain
\and JILA and Department of Astrophysical and Planetary Sciences,
University of Colorado, Boulder, CO 80309, USA
\and Department of Particle Physics \& Astrophysics, Weizmann Institute of Science, Rehovot 76100, Israel
}





\abstract{We use VLTI/GRAVITY near-infrared interferometry
  measurements of 8 bright, Type 1 AGN to study the size and structure
  of hot dust heated by the central engine. We partially resolve each
  source, and report Gaussian FWHM sizes in the range $0.3-0.8$
  milliarcseconds. In all but one object, we find no evidence for
  significant elongation or asymmetry (closure phases $\lesssim
  1^\circ$). The effective physical radius increases with bolometric luminosity as found from past reverberation and interferometry measurements. The measured sizes for Seyfert galaxies are systematically larger than for the two quasars in our sample when measured relative to the previously reported $R \sim L^{1/2}$ relationship explained by emission at the sublimation radius. This could be evidence of evolving near-infrared emission region structure as a function of central luminosity.}


\titlerunning{Resolved size of hot dust near AGN}
\authorrunning{GRAVITY Collaboration}

\maketitle




\section{Introduction} \label{sec:intro}

The spectral energy distributions (SEDs) of AGN show an excess in the
near-infrared (NIR) due to thermal emission from hot dust grains with
a color temperature $T_{\rm color} \simeq 1000-1500$K. The dust
luminosity and temperature is explained as reprocessing of emission
from the central accretion disk \citep[e.g.,][]{rieke1978}. In the
standard unification paradigm of AGN, dust distributed in a flattened,
large covering factor, parsec-scale ``torus" obscures the view of the
broad emission lines and central accretion disk for large viewing
angles \citep{antonucci1993,urry1995}. The NIR radiation might then
arise at the inner edge near the sublimation radius, where irradiation
from the center is strongest and dust temperature the highest. The
physical origin of such a structure remains unclear. Support by gas
pressure would require a sound speed far in excess of the Keplerian
speed at that distance from the central black hole. Alternatives
include radiation or magnetic pressure support, usually in the form of
an outflow. 

Mid-infrared interferometry presents a challenge to the torus paradigm. Many objects show an unresolved core, consistent with an origin in the inner part of the torus. At the same time, a significant fraction of the total flux may originate in the polar region on pc and larger scales, attributed to dusty outflows from the center \citep[e.g.,][]{lopezgonzaga2016,hoenig2017}. It remains unclear whether this material is producing significant obscuration of the central source. The first resolved image of hot dust found from GRAVITY observations of NGC 1068 shows a size consistent with that expected for the sublimation radius, but in a geometrically thin ring geometry (GRAVITY Collaboration: Pfuhl et al., submitted). An additional obscuring structure is required to explain the absence of broad emission lines.

Continuum reverberation experiments find correlated variability between the optical and NIR emission with a lag consistent with reprocessing. The inferred emission radius scales with luminosity as $R \sim L^{1/2}$ \citep{suganuma2006,koshida2014}, as expected if hot dust radiation peaks near the sublimation radius where the central engine radiation is weak enough for dust to survive \citep[e.g.,][]{barvainis1987}. The normalization of the relation is smaller than predicted, which may be due to the presence of large, graphite dust grains \citep{kishimoto2007}. 

The inferred sub-pc ($\lesssim$ milliarcsecond, mas) scales are too compact to be spatially resolved with single telescopes. NIR interferometry, mostly with the Keck Interferometer \citep{swain2003,kishimoto2009,kishimoto2011} as well as the VLT Interferometer instrument AMBER \citep{weigelt2012} have measured compact sizes associated with partially resolved sources. The size measurements are consistent with an origin at the sublimation radius as found from reverberation. 

The second generation VLTI instrument GRAVITY has vastly improved
sensitivity and coverage as a result of combining light from all 4 UT
telescopes, resulting in a 6 baseline array
\citep{gravityfirstlight}. We use data from our ongoing AGN observing
program to measure sizes for a sample of 8 of the brightest type 1
AGN, nearly doubling the sample for which NIR interferometry is
available. We describe the data acquisition and selection procedure
(\autoref{sec:data}) and fitting methods used to measure sizes from
both continuum and differential visibilities
(\autoref{sec:fitting}). All AGN are partially resolved, with
consistent results from both methods (\autoref{sec:results}). We find
similar angular sizes for objects of similar flux but spanning four orders of magnitude in luminosity, meaning that the physical size of the hot dust emission increases with luminosity. The two luminous quasars in our sample are more compact than the Seyfert 1s observed. All measured sizes are broadly consistent with the radius-luminosity relation determined using previous NIR interferometry measurements. We discuss the implications of our results in terms of dust emissivity, composition, and the relation of the hot dust emission to that of the broad emission line region (\autoref{sec:discuss}).

\begin{table*}
  \begin{tabular}{llccccccc}
  \hline
  \hline
  Source & Obs. Date & Int. Time (min) & Seeing (") & Coherence time
  (ms) & Strehl & H & K & V\\
  \hline
PDS 456 & 2018-08-26 & 65 & 0.51-0.85 & 4.8-8.3 & 0.03-0.14 & 11.3 & 10.2 & 14.5 \\
& 2018-08-27 & 40 & 0.40-0.50 & 4.0-6.2 & 0.18-0.23 & 11.7 & 10.7 & \\
NGC 3783 & 2018-01-07 & 80 & 0.40-0.70 & 5.7-10.9 & 0.03-0.07 & 10.8 &
9.8 & 13.4\\
& 2018-01-08 & 80 & 0.47-0.73 & 5.8-10.5 & 0.01-0.03 & 10.8 & 9.8 & \\
& 2018-05-31 & 105 & 0.38-0.61 & 2.5-4.8 & 0.10-0.27 & 10.6 & 9.8\\
& 2019-02-16 & 186 & 0.5-0.9 & 6.0-13.8 & 0.09-0.22 & 11.0 & 10.2\\
& 2019-03-31 & 120 & 0.4-0.6 & 3.2-7.0 & 0.03-0.20 & ... & 10.2\\
3C 273 & 2017-07-07 & 40 & 0.44-0.77 & 4.6-6.5 & 0.06-0.11 & 10.9 & ... & 12.9\\
& 2018-01-08 & 40 & 0.44-0.59 & 6.9-9.0 & 0.03-0.13 & 10.9 & 10.0 \\
& 2018-05-30 & 90 & 0.48-0.68 & 2.9-4.1 & 0.05-0.15 & 11.0 & 10.1\\
Mrk 509 & 2017-08-04 & 60 & 0.31-0.56 & 5.5-8.5 & 0.10-0.16 & 11.7 & ... & 13.5\\
& 2017-08-05 & 55 & 0.46-0.71 & 6.5-8.5 & 0.09-0.16 & 11.6 & 10.9 &\\
& 2017-08-08 & 30 & 0.29-0.48 & 9.0-10.8 & 0.23-0.28 & 11.6 & 10.7 & \\
& 2018-08-26 & 20 & 0.61-0.80 & 7.5-8.1 & 0.08-0.13 & 11.4 & 10.7 & \\
NGC 1365 & 2018-01-07 & 15 & 0.54-0.65 & 6.6-10.9 & 0.02-0.03 & 10.9 & 10.0 &  \\
& 2018-01-08 & 15 & 0.64-0.78 & 4.8-6.4 & 0.06-0.08 & 10.8 & ... &\\
3C 120 & 2018-11-20 & 20 & 0.35-0.56 & 4.1-5.7 & 0.05-0.13 & 11.4 & 11.6 & 15.1\\
IRAS 09149-6206 & 2018-11-20 & 65 & 0.64-1.03 & 3.7-4.0 & 0.07-0.13 & 10.9 & 9.8 & 12.3\\
& 2019-02-16 & 96 & 0.5-1.2 & 5.7-9.2 & 0.04-0.14 & 10.5 & 9.7 & \\
Mrk 335 & 2018-11-20 & 5 & 0.48 & 8.2 & 0.05 & 11.0 & 11.2 & 13.9\\
                  & 2019-07-15 & 50 & 0.5-0.9 & 1.8-3.8 & 0.03-0.08 & 10.9 & 11.0 &\\
    \end{tabular}
\caption{Observation epochs, total integration time on source, seeing
  and coherence time conditions reported by the DIMM, and Strehl
  ratios and H band aperture magnitudes obtained from the GRAVITY
  acquisition camera (50 mas FWHM aperture). The H and K band magnitudes were flux calibrated using Simbad H magnitudes of calibrator sources, and are only reported when a calibrator was observed. A typical uncertainty is 0.2 (H) and 0.3 (K) mag. The K magnitudes measured for 3C 120 and Mrk 335 are likely much more uncertain due to the very low Strehl ratios of those observations. The V magnitudes are taken from Simbad.\label{tab:1}}
\end{table*}

\section{Observations, Data Reduction, and Data Selection}\label{sec:data}

\subsection{Observations}
\label{sec:obs}

The main science goal of our GRAVITY AGN observing program is to spatially resolve the broad emission line region, which has recently been achieved with observations of the quasar 3C 273 \citep{gravity3c273}. Targets are selected as the brightest Type 1 AGN on the sky visible from the VLTI. For BLR science we require deep integrations and repeated observations over many nights, with less emphasis on observing calibrators.

In the past 2 years we have successfully observed 8 Type 1 AGN over 26
nights\footnote{Observations were done using the ESO Telescopes at the La Silla Paranal Observatory
  programme IDs 099.B-0606, 0100.B-0582, 0101.B-0255, 0102.B-0667,
  1103.B-0626.}. Details of the targets and observations are given in
Table \ref{tab:1}. For each observation, we first close the loop of
the MACAO visible AO systems on the visible AGN source with each of
the 4 UT telescopes. The AGN is then acquired on the GRAVITY
acquisition camera and we place both the GRAVITY fringe tracking (FT)
and science channel (SC) fibers on the AGN (observing on-axis), and
split the light between the two detectors. Once fringes are acquired,
we collect exposures with coherent integrations of 3.3ms on the FT and
30s on the medium spectral resolution SC detector ($R \sim 500$). We record a sequence of exposures of 10 SC DITs each, interrupted by occasional sky exposures or calibrator observations. Even the brightest AGN are relatively faint in V, resulting in a poor AO correction even in exceptional conditions. Fringe tracking signal-to-noise is generally low ($\lesssim 3$) at 300 Hz but remains stable due to the excellent performance of the GRAVITY fringe tracker \citep{lacour2019}.

\subsection{Data Reduction}

The data are reduced with the standard GRAVITY pipeline \citep{lapeyrere2014,gravityfirstlight} in two separate modes. For the continuum FT visibility data we use the default pipeline settings. The low signal-to-noise and reduced SC visibility amplitude (loss of coherence) often result in the pipeline flagging SC DITs or entire exposures. In analyzing those data, we have found a substantial improvement in performance by retaining all data independent of fringe tracker signal-to-noise or V-factor and then averaging all DITs by the inverse variance of their differential phases \citep{gravity3c273}. The differential phase is constructed by removing a mean and slope across the spectrum. We use a method where the mean and slope are computed separately for each spectral channel, excluding that channel itself from the measurement \citep{millour2008}, as implemented in the GRAVITY pipeline. This method improves our differential phase precision by $\simeq 10-20\%$.




\subsection{Acquisition Camera Images and Photometry}
\label{sec:acqcam}

We estimate the AGN H band magnitude using the GRAVITY acquisition camera. We fit a 2D Gaussian model to the image from each telescope in each exposure, and estimate the source flux as that integrated over the Gaussian model. We then flux calibrate using the same acquisition camera measurements of calibrator stars of known magnitude. The results are listed in Table \ref{tab:1}. From looking at the same source on successive nights, or using different calibrators within the same night, we estimate the uncertainty in these measurements as $0.2$ mag.


\begin{figure*}
\centering
\includegraphics[width=0.9\textwidth]{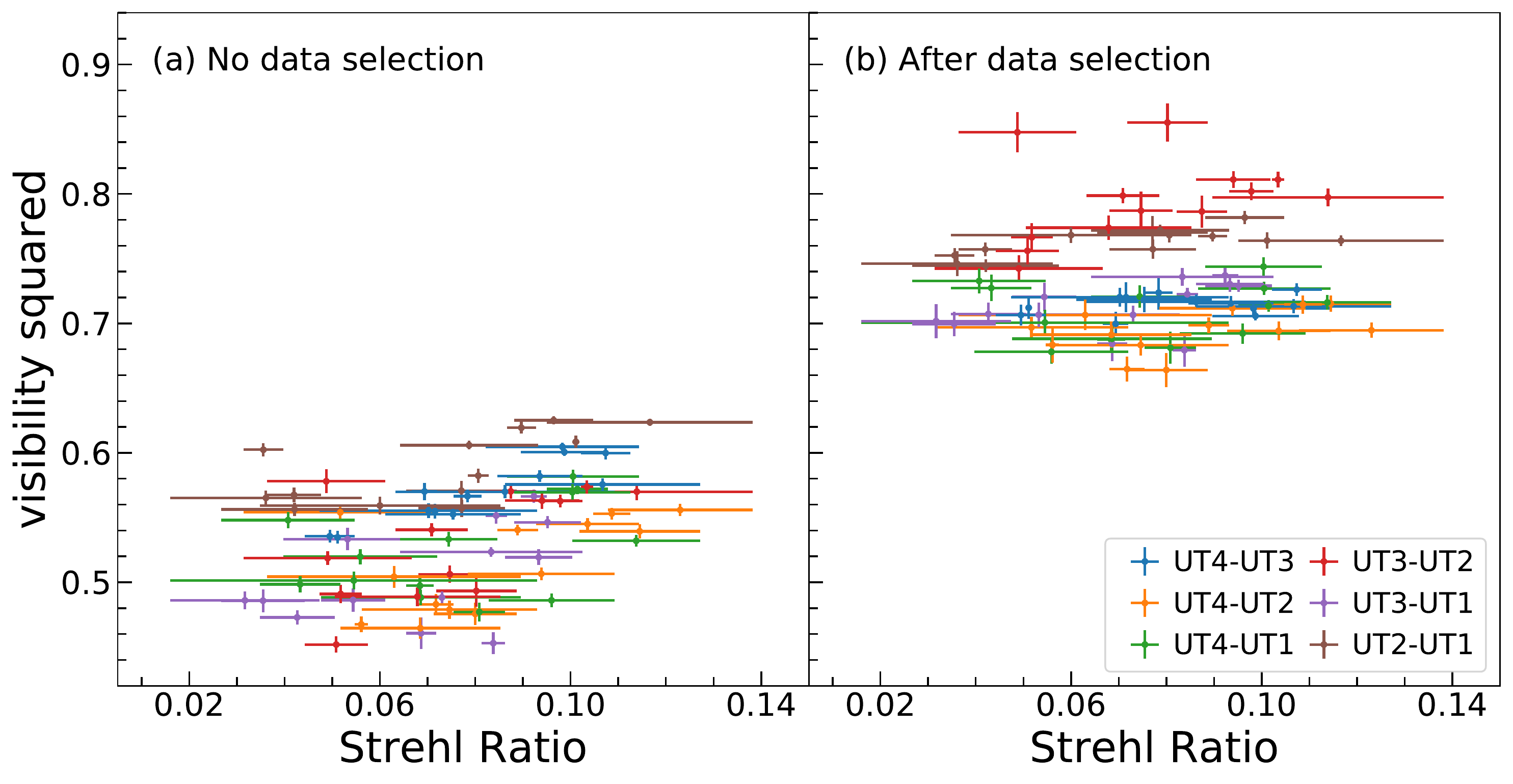}
\caption{Sample $V^2$ data of PDS~456 on 2018-08-27 before and 
after the data selection (group delay of $< 3 \mu$m). Without the data selection (a), the averaged visibility squared data are $\lesssim 0.65$. There is a trend that visibility squared increases as the Strehl ratio increases for each baseline (e.g., UT4--UT3 and UT4--UT2). In contrast, the visibility squared reaches $\simeq 0.8$ after the data
selection with no clear trend with Strehl ratio.}
\label{fig:sel}
\end{figure*}

\begin{figure*}
\centering
\begin{tabular}{cc}
\includegraphics[width=0.48\textwidth]{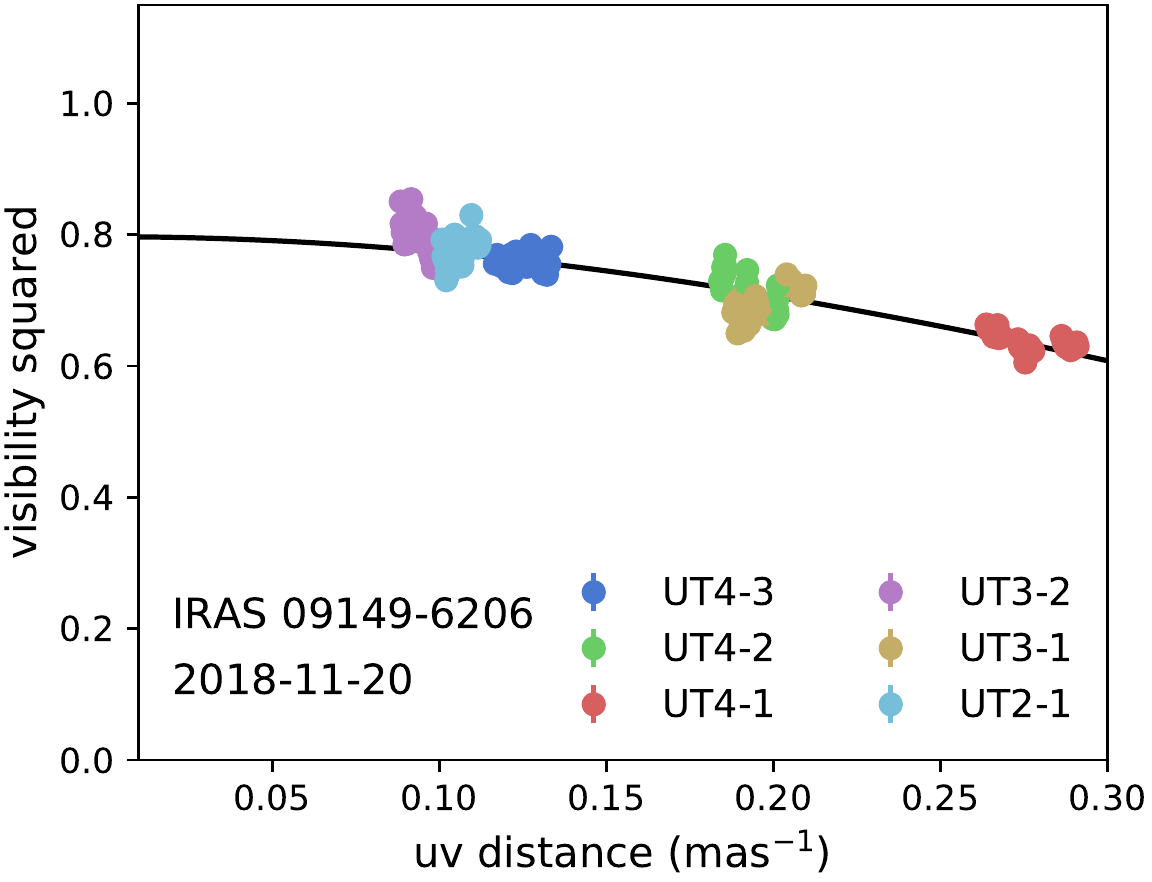} &
\includegraphics[width=0.48\textwidth]{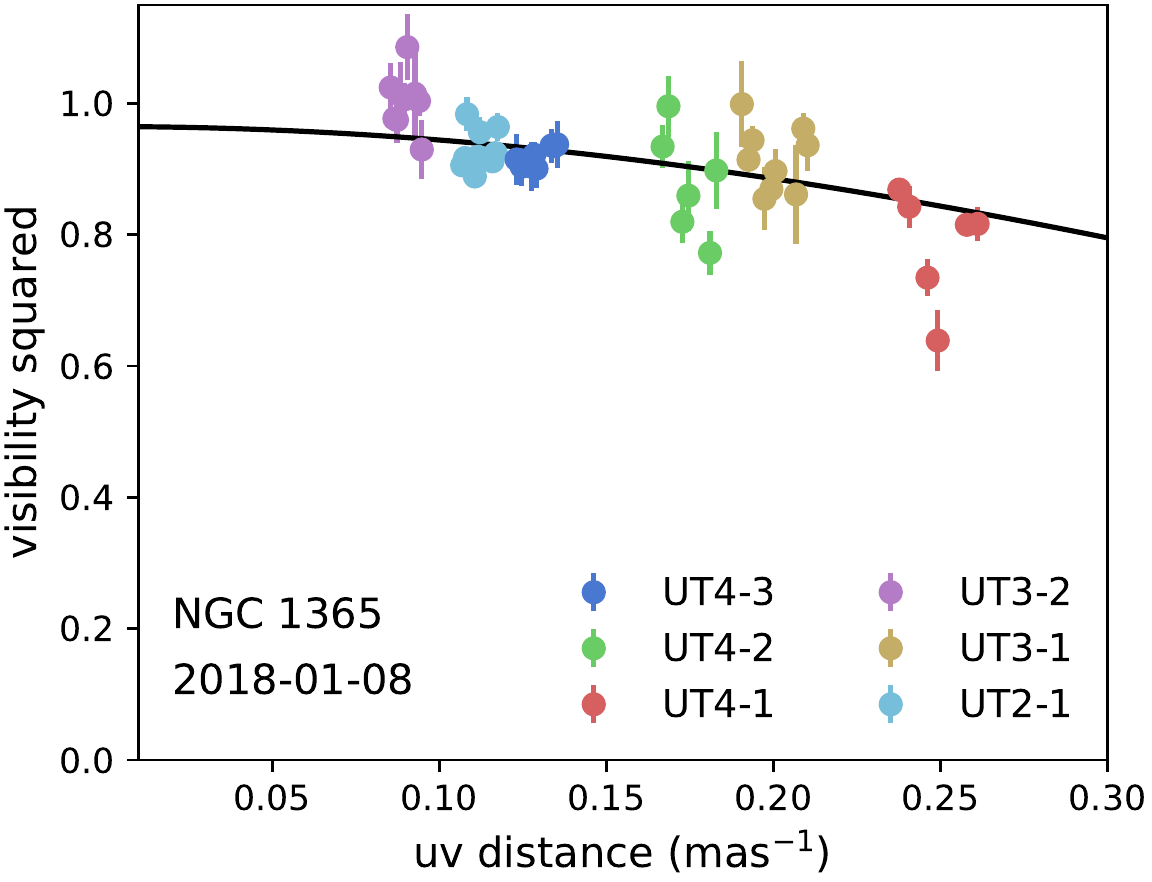}

\end{tabular}
\caption{Sample calibrated $V^2$ GRAVITY FT data as a function of
  baseline length for single epoch observations of two sources. The
  data pointed are color-coded by baseline. In both cases the sources
  are partially resolved, e.g. the visibilities fall with baseline
  length. For NGC 1365 on 2018-01-08, the $V^2$ approaches 1 at short
  baselines. For IRAS 09149-6206 on 2018-11-20, the short baselines
  remain significantly below $V^2=1$. The black lines are the best
  fitting 1D Gaussian models for each epoch.\label{fig:vis2ruvdata}}
\end{figure*}

\subsection{Data Selection}

The fringe tracker (FT) keeps the record of fringe measurement in
every DIT of 3.3~ms.  The group delay and phase delay are calculated
in real time to track the variation of the optical path differences of
the baselines. The fringe tracker stabilizes the fringe to allow long
integrations of the science channel. If the tracking source is bright,
most ($\gtrsim 90\%$) of the FT group delays are $\lesssim 3 \mu$m.  However, the group delay of a considerable fraction (e.g., $\gtrsim 50\%$) of the AGN fringe tracking data are usually much larger, 
resulting in considerable visibility loss. As an example in
\autoref{fig:sel}, if we average all of the FT data, the visibility is
much lower than unity.  The correlation of the visibility squared with
Strehl ratio implies that the visibility loss is likely affected by
the atmosphere. When we only select the DITs with group delay $<
3\,\mu$m (rejecting $50-80\%$ of DITs), the averaged visibility squared data are apparently improved. The correlation of visibility 
squared with Strehl ratio also vanishes.  Although there are other effects possibly still keeping the visibility of the shortest baseline (UT3--UT2) below unity, the data selection based on the group delay apparently helps to improve the data.


\section{Visibility model fitting}\label{sec:fitting}

The typical size scales of hot dust in Type 1 AGN are $\lesssim 1$ mas while the maximum VLTI/UT baseline resolution is $\simeq 3$ mas. As a result, all of our targets are only partially resolved. In that limit, all source models predict a similar trend of visibility amplitude (or $V^2$) with uv distance (spatial frequency). The observable property is then the characteristic size, related to the normalized second moment of the image \citep[e.g.,][]{lachaume2003,johnson2018}. We adopt Gaussian source models throughout, and use both FT and SC data to measure the characteristic size (FWHM) of the hot dust emission region. We report angular measurements as Gaussian FWHM to avoid assumptions about the hot dust emission, e.g. that it comes from a thin ring near the sublimation radius as expected for an obscuring torus.

\subsection{Continuum visibility fitting}

Sample FT continuum $V^2$ data for two sources are shown in  \autoref{fig:vis2ruvdata}, color-coded by baseline. In both cases the source is partially resolved, e.g. the visibilities fall with increasing baseline length. For NGC 1365, the visibility at zero baseline reaches unity, as it should for a model of a single compact source. In IRAS 09149-6206, it reaches only $V^2(0) \simeq 0.8$. The zero baseline visibility varies between sources but also between exposures and nights for the same target. We attribute this to a likely coherence loss, although some fraction could also result from extended nuclear K band continuum emission. To deal with this effect, we fit for the zero baseline visibility along with the source size in each exposure for each object over all nights. The function fit is then,

\begin{equation}
    V^2 = V_0^2 \exp{(-2 \pi^2 r_{\rm uv}^2 \sigma^2)},
  \end{equation}

  \noindent for zero baseline visibility $V_0$, size parameter
  $\sigma$ in radians, and $r_{\rm uv}$ the baseline length in units
  of the observed wavelength $\lambda$. We report sizes after
  converting $\sigma$ to FWHM measured in mas. Sample fits are shown
  as the solid lines in \autoref{fig:vis2ruvdata}, plotted against all data from one night for each source.

We determine average sizes and uncertainties from the median and rms scatter of all measured sizes over all exposures. To account for correlated systematic errors in calibration, possibly related to AO performance and seeing conditions, we choose an uncertainty in the mean reduced by $\sqrt{N_{\rm epochs}}$ (nights) rather than by the total number of exposures.


\begin{figure}
\includegraphics[width=0.48\textwidth]{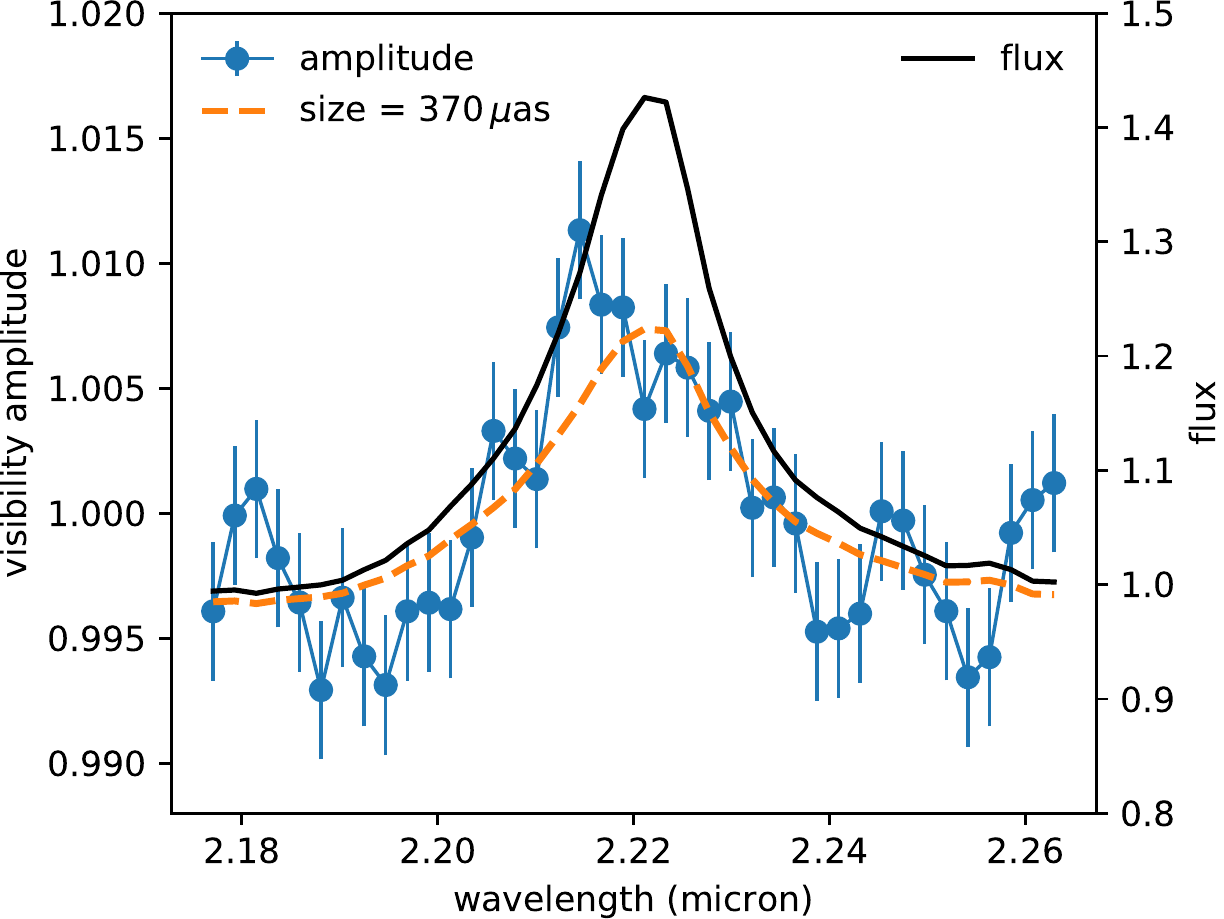}
\caption{Differential visibility amplitude vs. wavelength for PDS 456 from 2018-08-26 averaged over long baselines (blue circles and error bars) compared with the photometric flux. The orange curve shows a model of an unresolved BLR and a Gaussian continuum. The rise of the differential amplitude following the shape of the line is consistent with a partially resolved continuum, and its amplitude allows an independent measurement of its size.\label{fig:diffamp}}
\end{figure}

\begin{figure*}
  \centering
\begin{tabular}{cc}
\includegraphics[width=0.48\textwidth]{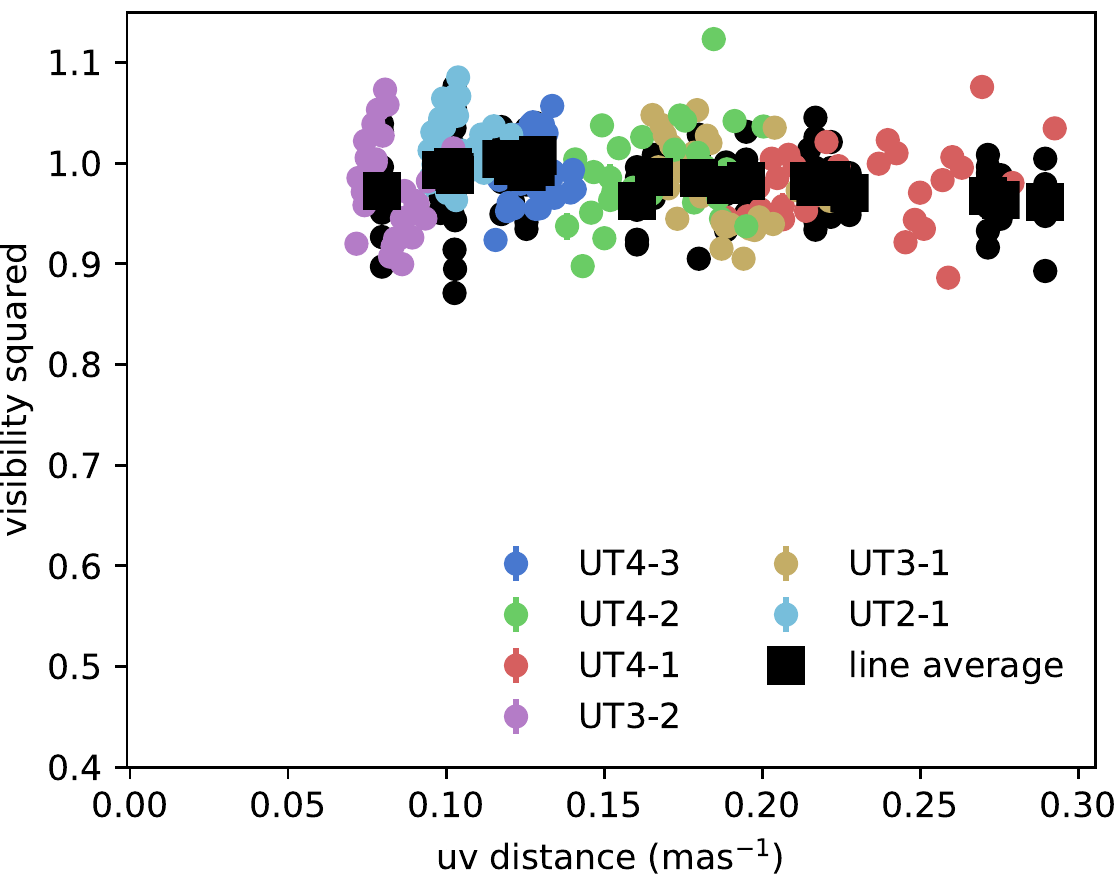}
\includegraphics[width=0.48\textwidth]{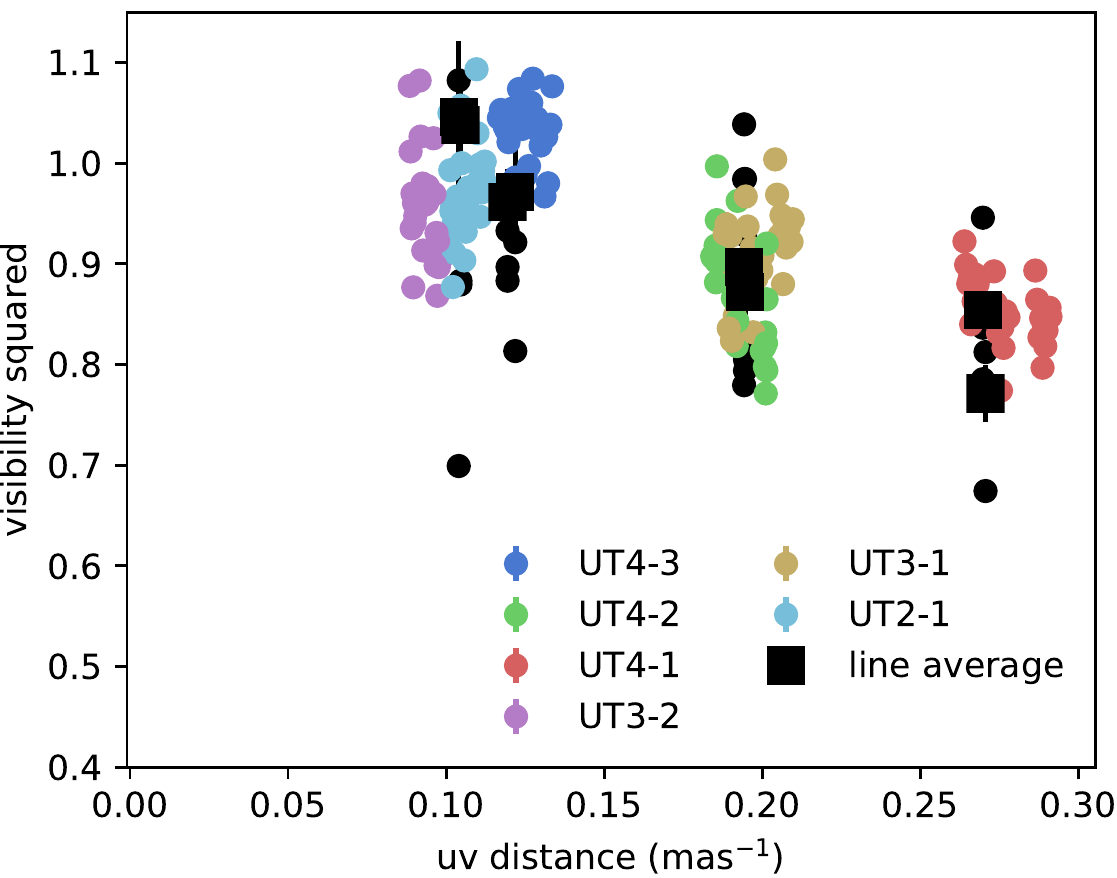}
\end{tabular}
\caption{Sample squared visibilities measured for FT (small circles colored by baseline) compared to those derived from SC differential amplitudes (black circles and averages as large squares) for 3C 273 (left) and IRAS 09149-6206 (right). The FT data have been divided by the best fitting zero baseline visibility, so that $V^2(0) = 1$. With that scaling, the two independent measurements are generally consistent. \label{fig:diffcompare}}
\end{figure*}

\subsection{Differential visibility amplitude analysis}

We also independently estimate hot dust continuum visibility amplitudes from the differential amplitude measured in the continuum and broad emission lines \citep[][low redshift Br $\gamma$ in Seyfert galaxies or Pa $\alpha$ for quasars at $z > 0.1$]{gravity3c273}. We normalize the visibility amplitude of each baseline and exposure to its median value, flatten slopes across the wavelength range by filtering out low frequency modes, and average over time with weights measured empirically in each exposure.

We construct the differential visibility by normalizing the
visibility amplitude of each baseline. At a spectral channel with line
flux $f$ relative to a continuum level of 1, this is

\begin{equation}\label{eq:diffvis}
v = \frac{1 + f V_l/V_c}{1+f},
\end{equation}

\noindent where $V_c$ and $V_l$ are the continuum and line
visibilities. The differential amplitude $v$ at the line depends on
the ratio $V_l/V_c$. In the partially resolved limit, that in turn
measures the quadrature size difference between the line and continuum
\citep{waisberg2017}. For a larger (smaller) spectral line emission
region, $|v|$ decreases (increases) at the line. We used these data
previously to show conclusively that the hot dust emission size is
larger than that of the Pa $\alpha$ BLR in 3C 273
\citep{gravity3c273}. The continuum visibility is

\begin{equation}\label{eq:vcvl}
\frac{V_c}{V_l} = \frac{f}{v (1+f) - 1}.
\end{equation}

We measure continuum visibilities $V_c$ from the data ($v$, $f$) by
assuming $V_l=1$, e.g. that the broad emission lines are
unresolved. The values $f$ are taken from the photometric flux
spectra, flattened to remove the instrument profile and averaged over
the 4 UT telescopes and all exposures. They are then normalized to the
continuum level. For 3C 273, we have found a BLR size from modeling differential phase data of $R_{\rm BLR} = 46 \pm 10 \mu$as \citep{gravity3c273}. That size corresponds to a visibility $V_l \simeq 0.996$ at a long VLTI baseline of $120$m, justifying our approximation. We note that in general the BLR is found from reverberation to be a factor of $\simeq 2-5$ smaller than the hot dust continuum \citep[e.g.,][]{koshida2014}. In that case, our approximation will tend to overestimate continuum visibilities, leading to a $\simeq 10-20\%$ underestimate of the continuum size.


These measurements require deep integrations, but do not suffer from systematic errors related to calibration or coherence loss. We measure $V_c$ from this method in 4 sources (3C 273, PDS 456, NGC 3783, IRAS 09149-6206). We expect $V_c$ to be independent over line spectral channel. From \autoref{eq:diffvis}, we then predict an amplitude signature $|v|$ which follows the shape of the spectral line, with a peak at the peak of the line where the contrast between line and continuum images is largest. An example is shown in \autoref{fig:diffamp} for PDS 456, where the visibility amplitude averaged over long baselines shows a significant increase at the spectral line, following the expected behavior.

Sample comparisons of our independent FT continuum and SC differential
measurements of the continuum visibility amplitude $V_c$ are shown in \autoref{fig:diffcompare}. They are generally consistent. Following our analysis of the FT data, we also perform single 1D Gaussian source fits to the SC differential data for each source to infer a characteristic FWHM size.

\begin{figure}
\includegraphics[width=0.48\textwidth]{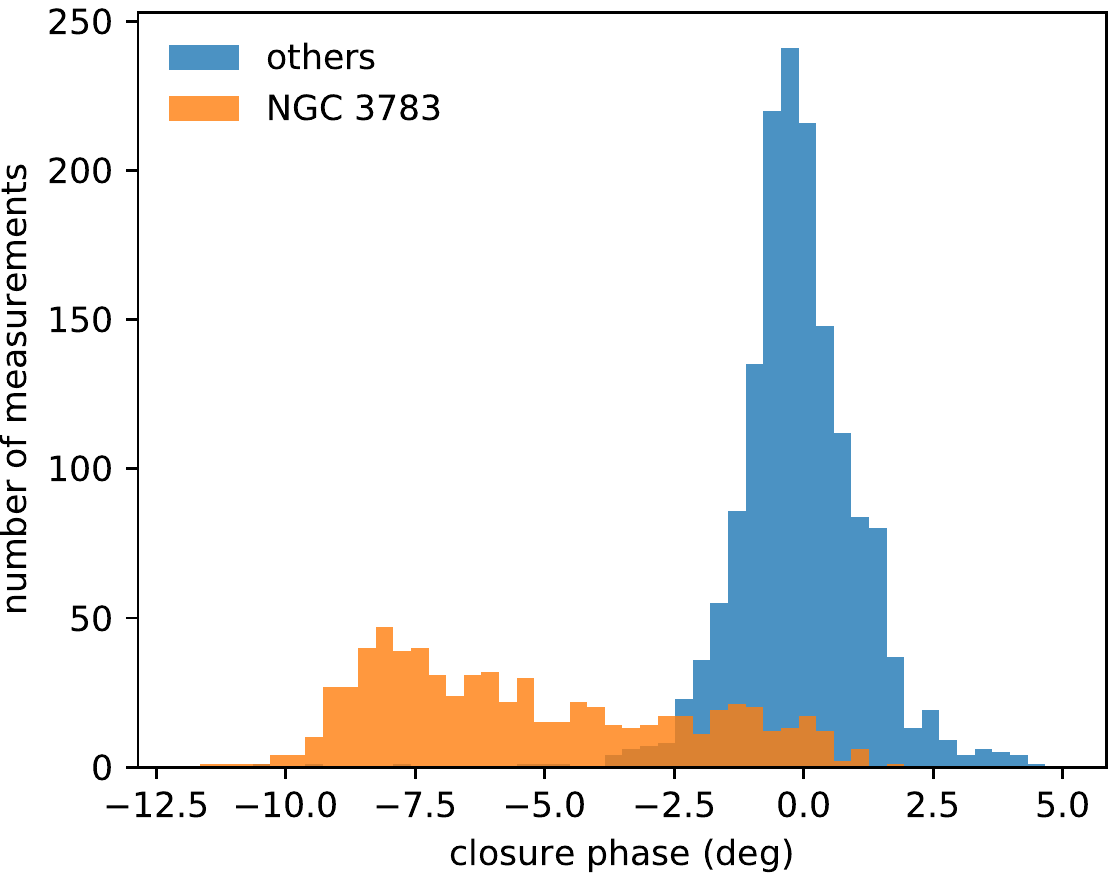}
\caption{Distributions of measured closure phases for NGC 3783 (orange) and other AGN targets with sufficient data (3C 273, PDS 456, Mrk 509, IRAS 09149-6206).\label{fig:t3phi}}
\end{figure}

\section{Results}
\label{sec:results}

\subsection{Elongation and asymmetry}

We do not find evidence for 2D (elongated) or asymmetric structure in
either the SC or FT data, except in NGC 3783, which will be analyzed
in more detail elsewhere. \autoref{fig:t3phi} shows distributions of
closure phases combined for 3C 273, PDS 456, IRAS 09149-6206, and Mrk
509 (our sources with the most and highest precision data). The
closure phase is formed by summing the visibility phase over baseline
triangles, and is immune to telescope-based phase errors. The
distributions show a typical closure phase rms of $\pm 1^\circ$ with a
median consistent with zero, indicating symmetric structure down to $<
0.1$ mas scales.\footnote{In the partially resolved limit the closure
  phase is to leading order $\propto (2 \pi u \cdot x)^3$ where $x$ is
  the size scale of the image \citep{lachaume2003}. This becomes extremely small for sizes $< 0.1$ mas regardless of intrinsic source structure.} The closure phases measured for NGC 3783 are shown as a separate histogram, are always $< 0^\circ$, and clearly indicative of asymmetry. For consistency with the other sources, we report size measurements for NGC 3783 from 1D Gaussian model fits, even though the model is inconsistent with the observed non-zero closure phases. 

\subsection{The hot dust is more extended than the BLR}

In all sources where we detect a differential amplitude signature, the amplitude increases at the line, showing robustly that the broad emission line region is more compact than the hot dust continuum. In the remaining targets we were limited by sensitivity, e.g. the size difference is not constraining. We confirm that the hot dust emission region is much larger than the BLR, as inferred from past reverberation and spectral measurements \citep[e.g.,][and references therein]{netzer2015}. In 3C 273 we also have an interferometric BLR radius measured from kinematic modeling of the detected velocity gradient in differential phase data \citep{gravity3c273}, which is a factor $\simeq 3$ smaller than that of the hot dust.

\begin{figure}
\includegraphics[width=0.48\textwidth]{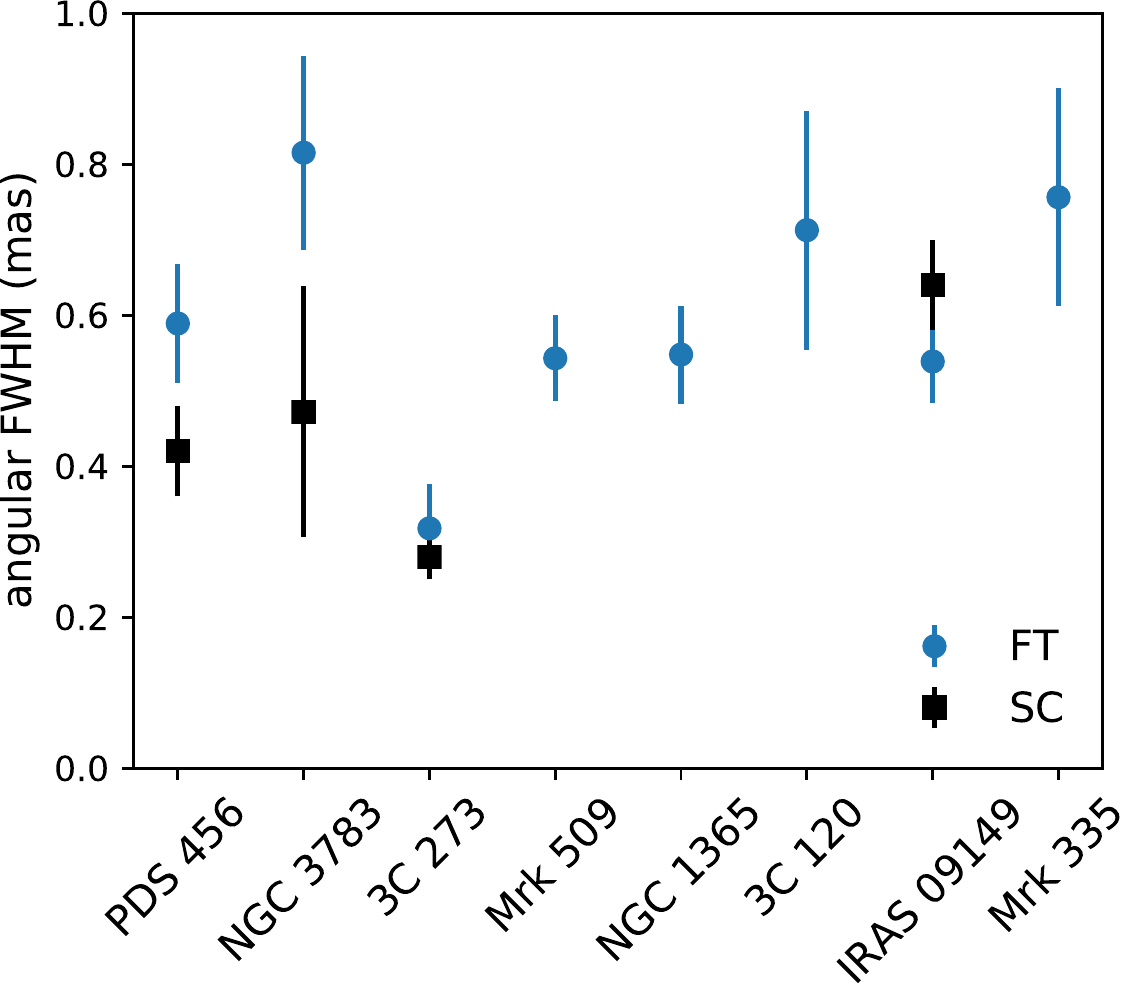}
\caption{Average source angular sizes (Gaussian FWHM) and their
  uncertainties measured from both continuum (FT) and spectral
  differential (SC) data. The latter are measured in deep
  integrations and are only possible for 4 objects so far.\label{fig:ang_size_compare}}
\end{figure}


\begin{table*}
\centering
	\caption{Source zero baseline visibilities, angular and
          physical size measurements, dust surface emissivities, and
          bolometric luminosities. Angular sizes are reported as
          Gaussian FWHM, while for consistency with the literature physical radii are measured for a thin ring model assuming a $20\%$ point source fraction.}
	\label{tab:sizes}
\begin{tabular}{lcccccc}
\hline
  \hline
  source & FT $V_0$ & FT size (mas) & SC size (mas) & radius (pc) &
                                                                  emissivity
  & $\log L_{\rm bol}$\\
  \hline
PDS 456 & $0.98 \pm 0.02$ & $0.59 \pm 0.08$ & $0.42 \pm 0.06$ & $1.342 \pm 0.179$ & $0.13 \pm 0.03$ & $47.00$\\
NGC 3783 & $0.86 \pm 0.05$ & $0.82 \pm 0.13$ & $0.47 \pm 0.17$ & $0.110 \pm 0.017$ & $0.11 \pm 0.03$ & $44.52$\\
3C 273 & $0.95 \pm 0.03$ & $0.32 \pm 0.06$ & $0.28 \pm 0.03$ & $0.567 \pm 0.106$ & $0.71 \pm 0.27$ & $46.64$\\
Mrk 509 & $0.86 \pm 0.01$ & $0.54 \pm 0.06$ & ... & $0.249 \pm 0.026$ & $0.13 \pm 0.03$ & $45.31$\\
NGC 1365 & $0.98 \pm 0.01$ & $0.55 \pm 0.06$ & ... & $0.032 \pm 0.004$ & $0.24 \pm 0.06$ & $42.96$\\
3C 120 & $0.93 \pm 0.03$ & $0.71 \pm 0.16$ & ... & $0.318 \pm 0.071$ & $0.03 \pm 0.01$ & $45.28$\\
IRAS 09149-6206 & $0.85 \pm 0.02$ & $0.54 \pm 0.05$ & $0.64 \pm 0.06$ & $0.405 \pm 0.041$ & $0.30 \pm 0.06$ & $45.29$\\
Mrk 335 & $0.92 \pm 0.02$ & $0.44 \pm 0.12$ & ... & $0.155 \pm 0.041$ & $0.14 \pm 0.07$ & $44.80$\\
\hline
\end{tabular}
\end{table*}

\subsection{Hot dust size measurements}

FWHM 1D Gaussian size measurements for each source are shown in
\autoref{fig:ang_size_compare} and listed in \autoref{tab:sizes} for
the FT and where available also the SC measurements. All sources are
partially resolved with sizes $\lesssim 1$ mas, and as compact as
$\simeq 0.3$ mas (3C 273). Such small sizes relative to the VLTI
interferometric beam are still detected due to the high sensitivity
and improved uv-coverage of GRAVITY. The sizes measured by the SC and
FT methods are generally consistent. The scatter in our size
measurements is dominated by systematic calibration errors, which
cause scatter in the sizes measured in individual exposures that are
much larger than expected by the signal-to-noise of the individual
$V^2$ measurements. This is reflected in the scatter of their zero
baseline visibilities $V_0$.

Past NIR interferometry measured sizes using a thin ring (delta function) intensity distribution rather than a Gaussian as we have done here. Keck Interferometer observations found a hot dust radius of the quasar 3C 273 of $0.25 \pm 0.1$ mas assuming a thin ring model \citep{kishimoto2011}, which is equivalent to a Gaussian HWHM of $0.21 \pm 0.09$ mas and consistent with our Gaussian HWHM result of $0.15 \pm 0.03$ mas. VLTI/AMBER observations found a thin ring radius of $0.74 \pm 0.23$ mas for NGC 3783 (Gaussian HWHM of $0.56 \pm 0.17$ mas), consistent with our result of $0.42 \pm 0.06$ mas. We note that the Keck observations used only 1 baseline (compared to our 6), while the VLTI/AMBER data used 3 baselines but at lower signal-to-noise. In analyzing the AMBER data, \citet{weigelt2012} further assumed a fixed zero baseline visibility of unity.

\subsection{BLR size estimates}

We have inferred continuum visibilities and sizes from SC differential
data by assuming a point source BLR ($V_l = 1$). Alternatively, we can
take the measured size from the FT data and use the SC differential
data to infer the BLR size. For 3C 273, this gives $R_{\rm BLR} = 75
\pm 65\,\mu$as, consistent with both GRAVITY modeling of the SC
differential phases \citep{gravity3c273} and various estimates from
reverberation mapping \citep{kaspi2000,peterson2004,zhang2019}. For
IRAS 09149-6206, the result is an upper limit of $R_{\rm BLR} \lesssim
100 \mu$as. For PDS 456 and NGC 3783, taken at face value we would
obtain large BLR sizes $R_{\rm BLR} = 200 \pm 70\,\mu$as and $R_{\rm
  BLR} = 320 \pm 100\,\mu$as. The uncertainties are large because the
SC differential amplitude is weakly sensitive to BLR size for such small angular sizes relative to the baseline resolution. We also note that any systematic offsets between our FT and SC results would likely change the results by a large amount compared to the statistical uncertainty. Still, in principle combining FT and SC data with sufficient sensitivity provides a model-independent measurement of BLR size that does not rely on adopting a kinematic model of the broad emission line profile.

\begin{figure*}
  \centering
\includegraphics[width=0.6\textwidth]{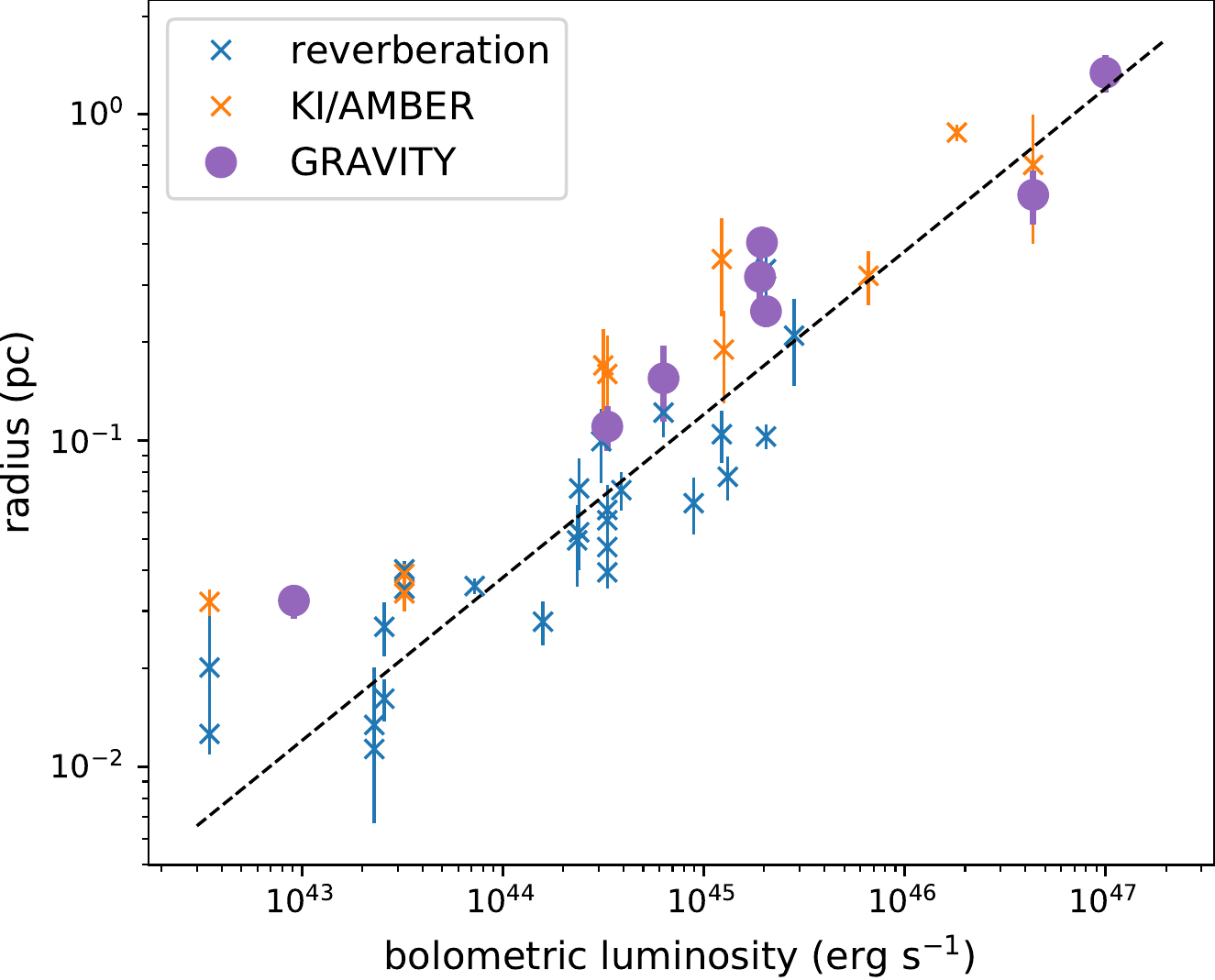}
\caption{Radius$-$luminosity relation for literature reverberation and interferometry size measurements (blue and orange x's) compared to our new GRAVITY measurements (filled circles). The dashed line is the best fitting $R \sim L^{1/2}$ fit to reverberation measurements \citep{suganuma2006,koshida2014}. We use FT measurements and our own bolometric luminosity estimates in all cases. The radii reported here for all interferometry measurements is that of a thin ring with an assumed unresolved point source flux fraction of $20\%$ (see \autoref{sec:rl} for details).\label{fig:rl}}
\end{figure*}

\begin{figure*}
  \centering
\includegraphics[width=0.6\textwidth]{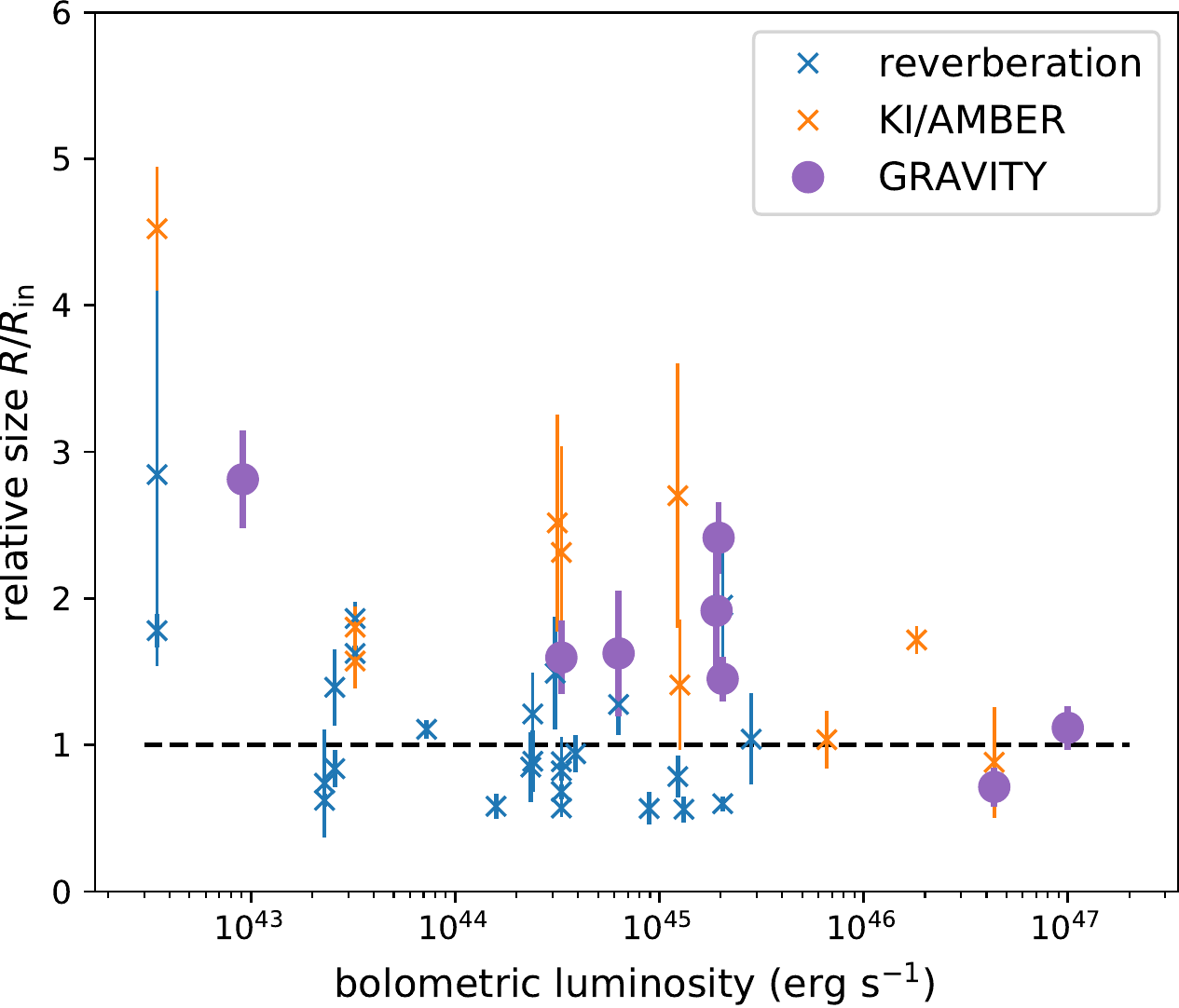}
\caption{Radius from GRAVITY and Keck/AMBER near-infrared interferometry, measured relative to the empirical radius-luminosity relation from reverberation mapping \citep{suganuma2006,koshida2014}. Reverberation sizes are shown for comparison and the error bars correspond to $1\sigma$ in all cases. Both interferometry data sets show a larger median size $R/R_{\rm in} > 1$ and decreasing relative size with increasing luminosity, particularly evident as $R/R_{\rm in} \lesssim 1$ for quasars.\label{fig:rin}}
\end{figure*}

\begin{table*}
\centering
	\caption{Reverberation lags, interferometric thin ring radii, and inferred geometric distances compared to fiducial ones given the AGN redshift. For NGC 3783, we have used the smaller size derived from differential SC data.}
	\label{tab:da}
\begin{tabular}{lcccc}
\hline
\hline
source & lag (lt-d) & $R_{\rm ring}$ (mas) & $D_A$ (Mpc) & $D_{A,\rm fiducial}$ (Mpc)\\
\hline
Mrk 335 & $148.9 \pm 24.1$ & $0.29 \pm 0.08$ & $90.0 \pm 27.8$ & 109.0\\
NGC 3783 & $73.0 \pm 14.0$ & $0.32 \pm 0.11$ & $39.8 \pm 15.9$ & 41.6\\
Mrk 509 & $126.8 \pm 11.0$ & $0.36 \pm 0.04$ & $62.3 \pm 8.5$ & 141.0\\
\hline
\end{tabular}
\end{table*}

\section{Discussion}
\label{sec:discuss}

We have measured the K-band continuum sizes of 8 bright Type 1 AGN using near-infrared interferometry with the VLTI instrument GRAVITY. We use continuum visibilities as measured by the GRAVITY fringe tracker $V^2$, and from the spectral differential visibility amplitudes. Each source is partially resolved, with FWHM sizes $\simeq 0.3-0.8$ mas smaller than the VLTI beam $\simeq 4$ mas. We find no evidence for significant elongation or asymmetry in 7 of 8 targets observed. We confirm that the BLR is more compact than the hot dust emission from direct measurements of the spectral differential visibility amplitude. Here we discuss the implications of our results for the size, structure, and physical properties of hot dust near AGN.





\subsection{Hot dust surface emissivity}

We measure the hot dust surface emissivity $\epsilon_\nu \equiv I_\nu / B_\nu$ (product of grain emissivity and surface filling factor) for objects in our sample, assuming a true dust temperature $T = 1500$ K (\autoref{tab:sizes}). This metric is equivalent to the surface brightness reported by \citet{kishimoto2011b}. Since our range of angular sizes is mostly consistent with past work, the dust surface emissivity values are as well $\epsilon_\nu \simeq 0.1-0.2$. The compact angular size of 3C 273 leads to a higher emissivity.

Since we do not resolve the emission region, our inferred values
correspond to averages over the effective VLTI beam, and are lower limits to the peak hot dust emissivity. The correction could be large, if a significant amount of the K band emission originates in a narrow ring \citep[][Gravity Collaboration et al., submitted]{kishimoto2009}. Such large surface emissivities close to blackbody emission can be expected if the dust emission is dominated by large grains $\ga0.2\,\mu$m as compared to a standard ISM grain size distribution with mean grain size $<0.1\,\mu$m. This is consistent with differential grain sublimation in the AGN vicinity \citep{hoenig2017}.

\subsection{The hot dust radius-luminosity relation}
\label{sec:rl}

Reverberation \citep{glass1992,suganuma2006,koshida2014} and near-infrared interferometry \citep{kishimoto2009,kishimoto2011,weigelt2012} observations have used hot dust continuum sizes to measure a radius-luminosity relation, $R \sim L^{1/2}$, consistent with the scaling for hot dust originating near the sublimation radius. At the same time, the normalization of that relation indicates a more compact than expected dust region, perhaps explained by the presence of large graphite grains \citep{kishimoto2007}. A relationship $R \sim L^{1/2}$ predicts equal angular size at constant observed flux \citep[e.g.,][]{elvis2002}. Here all objects are $K \simeq 10$, and their relatively constant angular sizes $\simeq 0.3-0.8$ mas over 4 orders of magnitude in bolometric luminosity show that the physical radius is indeed increasing with luminosity. 

Our observations nearly double the size of the near-infrared interferometry sample, and are measured with significantly improved uv-coverage and sensitivity. For comparison with past work, we measure physical radii by converting our Gaussian HWHM measurements to thin ring angular radii (divide by a factor of $\sqrt{\ln{2}} \simeq 0.8$). We then convert to a physical radius using the angular diameter distances to our targets. Significant point source contributions to the NIR source (from the accretion disk and/or jet) could cause us to underestimate the hot dust size. In the partially resolved limit, the source size corresponds to the normalized second moment of the image. Then the apparent source size for a point source plus extended model can be written,

\begin{eqnarray}
    \sigma_{\rm app}^2 &=& f_{\rm pt} \sigma_{\rm pt}^2 + (1-f_{\rm pt}) \sigma_{\rm true}^2,\\
    \frac{\sigma_{\rm app}}{\sigma_{\rm true}} &\simeq& \sqrt{1-f_{\rm pt}},
\end{eqnarray}

\noindent where $\sigma$ are normalized second moments for the point source component ($\sigma_{\rm pt}=0$), the total apparent source, and the true hot dust component. The fractional point source flux, $f_{\rm pt}$, varies per object. We adopt a typical constant value $f_{\rm pt} = 0.2$ in converting to physical radius \citep{kishimoto2007}, which results in a size increase of $\simeq 10\%$. The bolometric luminosities of both our objects and those from the literature are estimated uniformly as described in \autoref{apd:lbol}. The physical thin ring radii and bolometric luminosity values used are listed in \autoref{tab:sizes}.

\autoref{fig:rl} shows the resulting physical hot dust emission radius
vs. bolometric luminosity from past and our new measurements. Fitting
a power law $R \sim L_{\rm bol}^{1/2}$ to reverberation measurements
from the literature \citep{suganuma2006,kishimoto2007,koshida2014}, we
obtain the dashed line with a normalization $R_{44} = 0.038 \pm 0.02$
pc, where $R_{44}$ is the radius $c \Delta t$ at a bolometric
luminosity of $10^{44} \rm erg\,\rm s^{-1}$ and $\Delta t$ is the
measured lag between the $K$ and $V$ bands \citep[e.g.,][]{koshida2014}. Leaving the slope free, we
find a best fit with a flatter luminosity-dependence, $R \sim L_{\rm
  bol}^{0.40\pm0.04}$. The individual data points are plotted as blue
x's. Combining the Seyfert galaxies in our sample ($L_{\rm bol}
\lesssim 10^{46}\,\rm erg\,s^{-1}$) with past KI/AMBER results, we
find a similar best fitting radius-luminosity relation from NIR
interferometry as from reverberation. The two luminous quasars
($L_{\rm bol} > 10^{46}\,\ergs$, PDS 456 and
especially 3C 273) show smaller hot dust angular sizes than the other sources. Those sizes are robust, obtained with both SC and FT data, and precise (uncertainties $\lesssim 20\%$). The measurement for 3C 273 agrees with that reported by \citet{kishimoto2009}. We generally find that interferometry sizes are larger than those from reverberation at comparable bolometric luminosity, in agreement with past work \citep[e.g.,][]{kishimoto2011,koshida2014}.


\autoref{fig:rin} shows NIR interferometry radius measurements scaled to the $R-L$ relation from reverberation. The scatter between objects is larger than their errors, showing that there are differences in viewing geometry and/or physical structure (geometry or dust composition). At low luminosities, the interferometry sizes are systematically above the relation. Larger interferometric sizes might be expected as a result of an extended NIR emission region due to their weighting by total intensity, rather than by the response to a variable central source \citep[e.g.,][]{koratkar1991}. \citet{kishimoto2011b} argued that the reverberation value corresponds to the sublimation radius, $R_{\rm in}$, and that values $R/R_{\rm in} \simeq 1$ imply a hot dust emissivity falling rapidly away from that location. Larger values instead correspond to a shallower emissivity. In \autoref{fig:rin} we see that $R/R_{\rm in}$ falls with increasing bolometric luminosity, if with significant scatter. One explanation is a variable radial dust emission profile, which is sharply peaked at the sublimation radius for more luminous objects and slightly more extended at lower luminosity. Our data show more compact hot dust at high luminosity, and support this interpretation.


\subsection{Compact hot dust in quasars}

The two luminous sources also have $L/L_{\rm Edd} \sim 1$ and could have particularly high point source flux contributions. A bias by a factor of $2$, needed to bring 3C 273 within the observed range of other objects, requires a point source fraction $f_{\rm pt} = 3/4$, e.g. the observed K band emission needs to be dominated by contaminating emission. \citet{kishimoto2011} estimate an accretion disk contribution for 3C 273 of $f_{\rm pt} \simeq 0.3$, similar to our assumed $f_{\rm pt} = 0.2$. Non-thermal jet emission could also contribute in the NIR, as seen in rapidly variable K band light curves \citep[e.g.,][]{robson1993,mchardy2007,bewketubelete2018} corresponding to flaring events. Our observations of 3C 273 have produced consistent acquisition camera H band magnitudes and size measurements from July 2017 to May 2018. The acquisition camera H band flux density of $\simeq 40$ mJy is consistent with the quiescent state of 3C 273 with no evidence of flare contributions \citep[e.g.,][]{robson1993,soldi2008}. \citet{robson1993} estimate a jet contribution to the quiescent flux density of $\simeq 30$ mJy from extrapolating the power law spectral slope seen at millimeter wavelengths. \citet{soldi2008} find a significant accretion disk fraction but a much smaller jet contribution in SED fitting due to an imposed spectral cutoff to the synchrotron spectrum. \citet{kishimoto2011} estimate this contribution as $\lesssim 8$ mJy based on the low NIR polarization fraction and assuming an intrinsic $\simeq 10\%$ net polarization for the synchrotron component. These estimates produce a wide range of possible point source flux fractions from jet emission (assuming a small jet emission size), $f_{\rm jet} \simeq 0.1-0.4$. Combining the highest inferred values for both the accretion disk and jet contributions, the total point source contribution could plausibly be large enough to explain our observed compact size. A total point source contribution $\lesssim 0.5$ seems more likely based on the clear thermal dust bump in the NIR SED \citep{soldi2008}, and would increase our hot dust size by $\lesssim 20\%$. PDS 456 also shows a compact size  and is radio quiet. We conclude that contaminating emission can contribute to the compact sizes we measure, particularly for 3C 273, but probably does not explain our finding of $R/R_{\rm in} \lesssim 1$ in two luminous quasars.

A further bias comes from errors in the bolometric luminosity estimates. Even random errors in $L_{\rm bol}$ will tend to produce an anti-correlation of $R/R_{\rm in}$ and $L_{\rm bol}$, because $R_{\rm in} \propto L^{1/2}$. By simulating many random data sets, we find that the degree of anti-correlation observed in GRAVITY and KI/AMBER data could be explained if the true luminosity errors are $\simeq 0.6$ dex. Our measured errors by comparing various methods appear to be somewhat smaller $\simeq 0.3-0.5$ dex (\autoref{apd:lbol}). Still this bias warrants caution in interpreting the results in terms of varying hot dust structure with source luminosity.

\subsection{Geometric distance estimates}

Past measurements of the hot dust radius from reverberation have been made for NGC 3783 \citep{lira2011}, Mrk 335, and Mrk 509 \citep{koshida2014}. Geometric estimates of the angular diameter distance from combining time lags with interferometric sizes are given in \autoref{tab:da}. For those estimates we have followed \citet{koshida2014} in assuming $R = c \Delta t = D_A \theta_{\rm ring}$ where $\Delta t$ is the reverberation lag and $\theta_{\rm ring}$ is the angular radius of a thin ring model with a 20\% unresolved point source contribution. We find consistent angular diameter distances to the fiducial values given the redshift for Mrk 335, as well as for NGC 3783 if we use its differential SC angular size. The GRAVITY inferred physical radius of Mrk 509 is a factor $\gtrsim 2$ larger than from reverberation, given its fiducial angular diameter distance. A similar discrepancy is found using the continuum FT size of NGC 3783, although that object clearly shows complex structure beyond a partially resolved thin ring. Understanding the discrepancy and the physical structure of the NIR emission region will be important for efforts to measure geometric distances by combining interferometric and reverberation sizes \citep{hoenig2014}.

\section{Summary}

We have reported near-infrared interferometry measurements of 8 AGN
made with the second generation VLTI instrument GRAVITY. In all cases,
we partially resolve the continuum hot dust emission region and can
measure its size. In the 4 objects with sufficiently deep
integrations, we use a new spectral differential method to show
definitively that the hot dust continuum is much larger than the
low-ionization Pa $\alpha$ or Br $\gamma$ broad emission line
region. In 7/8 objects, we see no clear evidence for elongation or
asymmetry and in particular constrain the closure phases to $\lesssim
1^\circ$. The hot dust continuum sizes span $0.3-0.8$ mas, mostly
consistent with past interferometry measurements with the Keck
Interferometer and the VLTI instrument AMBER. Our objects span 4
orders of magnitude in bolometric luminosity with similar $V$ magnitude. Their roughly constant angular size implies increasing physical radius with luminosity, consistent with past reverberation and interferometry measurements. We find that at low luminosity, the hot dust sizes are systematically larger than those from reverberation, potentially related to emission from an extended region. At high luminosity, the two quasars in our sample are compact, with sizes consistent with sharply peaked emission at the sublimation radius. Accretion disk and/or jet contributions to the flux and biases from errors in estimated bolometric luminosities are probably not sufficient to fully explain this finding.


\begin{acknowledgements}
J.D., M.B., and A.J.R. were supported by a Sofja Kovalevskaja award from the Alexander
von Humboldt foundation and in part by NSF grant AST 1909711. SFH
acknowledges support by the EU Horizon 2020 framework programme via
the ERC Starting Grant DUST-IN-THE-WIND (ERC-2015-StG-
677117). M.K. acknowledges support from JSPS (16H05731). F.E. and O.P. acknowledge support from ERC synergy grant No. 610058 (Black-HoleCam). A.A. and P.G. acknowledge funding from Funda\c{c}\~ao para a Ci\^encia e a Tecnologia through grants UID/FIS/00099/2013 
and SFRH/BSAB/142940/2018. P.O.P acknowledges support from the Programme National Hautes Energies (PNHE) of the Centre National de la Recherche Scientifique (CNRS).
  \end{acknowledgements}

\appendix

\section{Bolometric luminosity}
\label{apd:lbol}

\begin{figure*}
\centering
\includegraphics[width=0.7\textwidth]{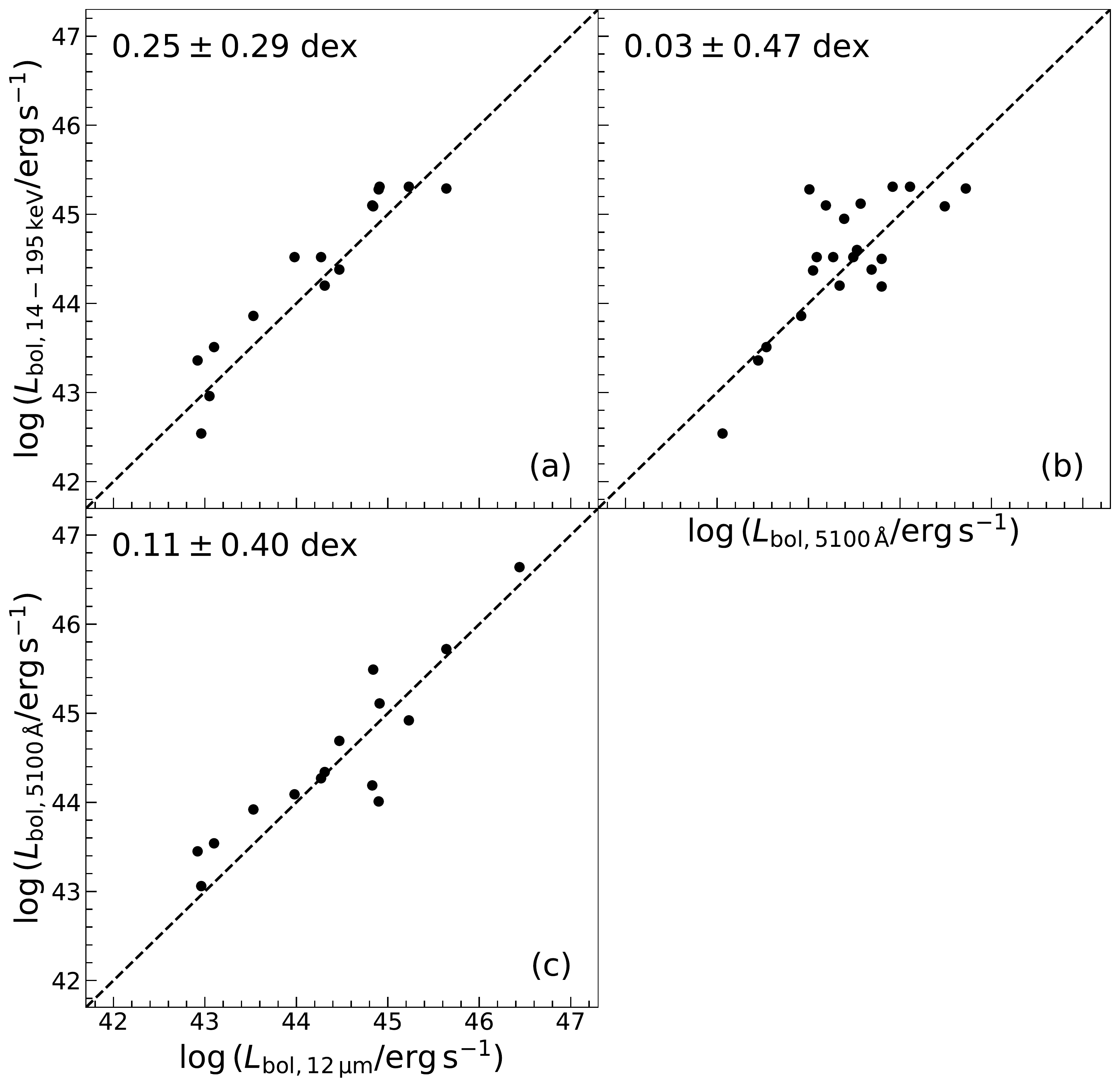}
\caption{Comparison of AGN bolometric luminosities derived using the different methods listed in Table \ref{tab:bol}.  The median and standard deviation of the 
difference ($y - x$) are indicated in the upper left corner of each panel. The dashed lines indicate the one-to-one relation.  (a) The 14--195 keV and 
12 $\mu$m results show a possible offset, comparable to their scatter ($\simeq 0.3$ dex).  The 5100 \AA\ bolometric luminosity tends 
to be better consistent with both 14--195 keV (b) and 12 $\mu$m (c), 
although the scatters of both relations are considerable.  Considering the 
uncertainty of the bolometric correction is large ($\sim 0.5$ dex), the 
different methods are relatively well consistent with each other.
}
\label{fig:bol}
\end{figure*}

\input{tab_lbol.tex}

The bolometric luminosity of the AGNs are estimated mainly with X-ray 
measurements.  Unless noted in particular, we take the 14--195 keV flux 
from the BAT AGN Spectroscopic Survey \citep{Koss2017ApJ}, taken directly from the 70-month Swift-BAT survey \citep{baumgartner2013}.  The hard X-ray emission is 
less contaminated by the host galaxy and less affected by the absorption.  
We adopt the luminosity dependent bolometric correction relation provided 
by \cite{Winter2012ApJ}, 
\begin{equation}
\log\, \left( \frac{L_{\mbox{\scriptsize bol, 14--195 keV}}}{\ergs} \right) 
= 1.1157\, \log\, \left( \frac{L_{\mbox{\scriptsize 14--195 keV}}}{\ergs} 
\right) - 4.2280.
\end{equation}
We also collect their optical (5100 \AA) continuum from BAT catalog and 
12 $\mu$m nuclear emission from \cite{Asmus2014MNRAS} to confirm the 
consistency of different methods to derive the bolometric luminosity.  
Following the 5100 \AA\ bolometric correction of \cite{Trakhtenbrot2017MNRAS},
we use,
\begin{equation}
\log\,\left( \frac{L_{\mbox{\scriptsize bol, 5100 \AA}}}{\ergs} \right) 
= 0.916\, \log\,\left( \frac{\nu L_\nu(\mbox{5100 \AA})}{\ergs} \right) 
+ 4.596.
\end{equation}
For the MIR data, we first convert the 12 $\mu$m flux to 2--10 keV 
intrinsic flux according to \cite{Asmus2011AA}, 
\begin{equation}
\log\,\left( \frac{f_{\mbox{\scriptsize 2--10 keV}}} {\mathrm{mJy}} \right) 
= 0.89\, \log\, \left( \frac{f_{\mathrm \scriptsize 12\,\mu \rm m}}{\mathrm{mJy}}
\right) - 12.81.
\end{equation}
Then we convert 2--10 keV luminosity to 14--195 keV luminosity according to 
\cite{Winter2009ApJ}, 
\begin{equation}\label{eq:210}
\log\,\left( \frac{L_{\mbox{\scriptsize 14--195 keV}}}{\ergs} \right) = 
0.94\, \log\, \left( \frac{L_{\mbox{\scriptsize 2--10 keV}}}{\ergs} \right) 
+ 2.91,
\end{equation}
and rely on the 14--195 keV bolometric correction \citep{Winter2012ApJ} to 
obtain the final bolometric luminosity. The latter step is to maintain consistency with our X-ray bolometric luminosity.  The intrinsic scatter of 
the relation in Equation (\ref{eq:210}) may introduce additional uncertainty. 
However, we believe this is not as important considering the large uncertainty 
of the bolometric correction (see below).  We compare the bolometric 
luminosities derived from the three methods in Figure \ref{fig:bol}.  The 
scatters of the relations are $\sim 0.3$--0.5 dex, which is likely reflecting 
the uncertainty of the bolometric correction methods at a single frequency.  
Considering the source variability and the difference of their intrinsic SEDs, 
the uncertainty of our bolometric correction is hardly below $\sim 0.5$ dex (see \citealt{Netzer2019MNRAS} and discussion therein).  Therefore, we conclude 
that the three methods produce mostly consistent results.  
A few objects are noted individually as follows, including the four objects 
(PDS~456, Mrk~231, IRAS~13349+2438, PG~1202+281) which are not available in the BAT 
AGN sample.  

\noindent
\textit{Mrk~335}: It is reported extremely variable in X-ray with amplitude 
a factor of $\sim 10$ and timescale tens of days \citep{Grupe2012ApJS}.  
Therefore, we prefer to adopt the 5100 \AA\ bolometric luminosity, which is 
consistent with the value reported by \cite{Woo2002ApJ}, who calculated the 
bolometric luminosity by integrating the observed flux over the SED.

\noindent
\textit{3C~120}: The bolometric luminosity derived from 5100 \AA\ luminosity 
is about 1 order of magnitude lower than those with other methods.  This is 
possibly due to variability.  However, we note that the 5100 \AA\ luminosity 
reported by \cite{Vestergaard2006ApJ} is $\sim 0.9$ dex higher than that of the
BAT AGN Spectroscopic Survey. The former provides a bolometric luminosity more consistent with our other methods.  The X-ray derived bolometric 
luminosity is also consistent with that integrated over the SED from \cite{Woo2002ApJ}.

\noindent
\textit{3C~273}: The jet emission is likely contributing significantly to 
the X-ray luminosity of 3C 273 \citep{Dermer1997ApJS,Vasudevan2007MNRAS}.  
Therefore, we discard the bolometric luminosities derived from X-ray data
and adopt the value based on 5100 \AA, which is found to be consistent with 
that based on 12 $\mu$m data.

\noindent
\textit{PDS~456}: This source is very luminous in UV/optical.  Its 5100 
\AA\ luminosity is $\sim 2\times 10^{46}\,\ergs$ 
\citep{Simpson1999MNRAS,Reeves2009ApJ}, corresponding to 
$L_{\mbox{\scriptsize bol, 5100 \AA}} = 1.0 \times 10^{47}\,\ergs$.  This 
is very well consistent with the bolometric luminosity derived from UV--IR 
SED \citep{Simpson1999MNRAS}.  PDS~456 is extremely 
variable in X-ray \citep{Reeves2002MNRAS} and the observed 2--10 keV 
luminosity is only 0.2\% of the bolometric luminosity \citep{Reeves2009ApJ}.
Therefore, we do not quote the bolometric luminosity estimated from X-rays.

\noindent
\textit{Mrk~231}: Recent \textit{NuSTAR} measurements reveal Mrk 231 to 
be intrinsically X-ray weak \citep{Teng2014ApJ}.  Therefore, X-ray measurements 
are not suitable for deriving the bolometric luminosity for this target.  We 
adopt the 12 $\mu$m measurement from \cite{LopezRodriguez2017MNRAS} and obtain 
the bolometric luminosity $\sim 6.6\times 10^{45}\,\ergs$, which is consistent 
with the bolometric luminosity, $\mbox{4--5} \times 10^{45}\,\ergs$, derived 
from the decomposition of IR SED \citep{Farrah2003MNRAS,Teng2014ApJ}.

\noindent
\textit{IRAS~13349+2438}:  We derive the bolometric luminosity of 
IRAS~13349+2438, $\sim 1.8 \times 10^{46}\,\ergs$, based on 12 $\mu$m 
data.  This is within a factor of 2 consistent with the estimates based 
on quasar SEDs \citep{Beichman1986ApJ,Lee2013MNRAS}.

\noindent
\textit{IC~4329A}: The bolometric luminosity derived from X-ray data is 
$\sim 0.9$ dex higher than that from 5100 \AA.  This is likely reflecting the 
uncertainty of the different bolometric luminosity corrections.  The 5100 
\AA\ luminosity of BAT catalog is consistent with other measurements (e.g., 
\citealt{Bentz2009ApJ}).  Meanwhile, the X-ray bolometric luminosity is 
consistent with the values derived from the AGN SEDs \citep{Woo2002ApJ,
Vasudevan2010MNRAS}.  Therefore, we still adopt the X-ray derived 
bolometric luminosity in our analysis.  It is worth noting that, with the 
bolometric correction theoretically calculated by \cite{Netzer2019MNRAS}, 
the bolometric luminosity based on 5100 \AA\ luminosity is $\sim 0.4$ dex 
higher than what we quote, and as a result is more consistent with that derived from X-rays.

\noindent
\textit{PG~1202+281}: We collect the 5100 \AA\ data from 
\cite{Vestergaard2006ApJ}.  The derived the bolometric luminosity is well 
consistent with that derived from NIR-to-X-ray SED \citep{Runnoe2012MNRAS}.

\noindent
\textit{Mrk 744}: Our X-ray derived bolometric luminosity is well consistent 
with that reported in \cite{Woo2002ApJ} interpolating the AGN SED.

\textit{Mrk 110}: We collect the 5100 \AA\ luminosity from
\cite{Vestergaard2006ApJ}.

\textit{IRAS 03450+0055}: The 5100 \AA\ luminosity comes from
\cite{Afanasiev2019MNRAS}.

%

\vspace{5mm}







\bibliography{bib}



\end{document}

%% file: tab_lbol.tex
\begin{table*}
  \centering
  \begin{tabular}{lrcccc}
    \hline
    Source & $D_L$ (Mpc) & $\log\,L_{\mbox{\scriptsize bol, 14--195 keV}}$ &
                                                                       $\log\,L_\mathrm{bol,
                                                                       12\,\mu
                                                                             \rm
                                                                             m}$
    & $\log\,L_{\mbox{\scriptsize bol, 5100 \AA}}$ &
                                                     $\log\,L_\mathrm{bol,
                                                     use}$\\
    \hline
    \hline
           NGC 1365 &  16.6 &       42.96 &       43.05 &      ... & 42.96 \\
           NGC 3783 &  47.8 &       44.52 &       44.27 &        44.27 & 44.52 \\
            Mrk 335 & 113.6 &       44.19 &     ... &        44.80 & 44.80 \\
             3C 120 &  144.9 &       45.28 &       44.90 &        44.01 & 45.28 \\
            Mrk 509 &  149.4 &       45.31 &       44.91 &        45.11 & 45.31 \\
  IRAS 09149$-$6206 & 254.7 &       45.29 &       45.64 &        45.72 & 45.29 \\
             3C 273 & 755.0 &     ... &       46.44 &        46.64 & 46.64 \\
            PDS 456 & 893.2 &     ... &     ... &        47.00 & 47.00 \\
           NGC 4051 &  13.4 &       42.54 &       42.96 &        43.06 & 42.54 \\
           NGC 4151 &   9.9 &       43.51 &       43.10 &        43.54 & 43.51 \\
            Mrk 231 & 185.6 &     ... &       45.82 &      ... & 45.82 \\
    IRAS 13349+2438 & 499.8 &     ... &       46.26 &      ... & 46.26 \\
              Mrk 6 &  82.6 &       44.50 &     ... &        44.80 & 44.50 \\
            Ark 120 & 140.4 &       45.09 &       44.84 &        45.49 & 45.09 \\
           IC 4329A &  69.4 &       45.10 &       44.83 &        44.19 & 45.10 \\
           NGC 3227 &  18.8 &       43.36 &       42.92 &        43.45 & 43.36 \\
           NGC 7469 &  69.4 &       44.38 &       44.47 &        44.69 & 44.38 \\
        PG 1202+281 & 791.8 &     ... &     ... &        45.45 & 45.45 \\
          Fairall 9 & 203.9 &       45.31 &       45.23 &        44.92 & 45.31 \\
            Mrk 744 &  50.6 &       43.41 &     ... &      ... & 43.41 \\
           NGC 5548 &  73.8 &       44.52 &       43.98 &        44.09 & 44.52 \\
            Mrk 590 & 118.0 &       44.20 &       44.31 &        44.34 & 44.20 \\
 MCG +08$-$11$-$011 &  87.0 &       44.95 &     ... &        44.39 & 44.95 \\
             Mrk 79 &  95.8 &       44.52 &     ... &        44.49 & 44.52 \\
            Mrk 110 & 158.4 &       45.12 &     ... &        44.57 & 45.12 \\
           NGC 3516 &   51.5 &       44.37 &     ... &        44.05 & 44.37 \\
           NGC 4593 &  34.7 &       43.86 &       43.53 &        43.92 & 43.86 \\
            Mrk 817 & 138.2 &       44.60 &     ... &        44.53 & 44.60 \\
    IRAS 03450+0055 & 135.9 &     ... &     ... &        44.49 & 44.49 \\
\end{tabular}
\caption{AGN distances and bolometric luminosities. All luminosities
  are in units of \ergs. (1) Source name.  
(2) Luminosity distance, calculated with $H_0=70\,\kms$, 
$\Omega_\mathrm{\Lambda}=0.7$, and $\Omega_\mathrm{M}=0.3$. 
(3) Bolometric luminosity derived from 14--195 keV data. The X-ray data
come from BAT AGN Spectroscopic Survey \citep{baumgartner2013,Koss2017ApJ}. 
(4) Bolometric luminosity derived from 12 $\mu$m data.  The MIR data 
mainly come from \cite{Asmus2014MNRAS}, except that of Mrk 231 (see \autoref{apd:lbol}).  
(5) Bolometric luminosity derived from optical 5100 \AA\ continuum data.  
The optical data mainly come from BAT AGN catalog, except for those of 
PDS~456 and PG~1202+281 (see \autoref{apd:lbol}).  
(6) The bolometric luminosity that we adopt for the $R$--$L$ relation.
The 14--195 keV bolometric luminosities are preferred, unless the objects
are noted individually. \label{tab:bol} 
}
\end{table*}